# Unraveling the deterministic effect of the solid-state diffusion energy barrier for charge carrier on the self-discharge of supercapacitors

*Xiaohui Yan,[+] Yue He,[+] Xuncheng Liu, Siqi Jing, Jiajian Guan, Wei Gao, Sudip Ray, Yige Xiong, Taibai Li, Xiang Ge**

*Corresponding Author

Department of Materials and Metallurgy, Guizhou University, Guiyang, Guizhou 550025, China

E-mail: xge@gzu.edu.cn

**Abstract**

The further development of fast electrochemical devices is hindered by self-discharge. Current strategies for suppressing self-discharge are mainly focused on the extrinsic and general mechanisms including faradaic reactions, charge redistribution, and ohmic leakage. However, the self-discharge process is still severe for conventional supercapacitors. Herein, we unravel the deterministic effect of solid-state diffusion energy barrier by constructing conjugately configured supercapacitors based on pairs of pre-lithiated niobium oxides with similar intercalation pseudocapacitive process but different phases. This device works with a single type of charge carrier while materials with various diffusion barriers can be implanted, thus serving as an ideal platform to illustrate the influence of the diffusion barrier. The results show that the comprehensive effect of solid-state diffusion energy barrier and extrinsic effects drives the self-discharge process. Noteworthy, the diffusion barrier presents with an exponential form, which governs the self-discharge of supercapacitors. This work is expected to unravel the deterministic effect of the solid-state diffusion energy barrier and provide a general guidance for suppressing self-discharge for supercapacitors.

# TOC GRAPHICS

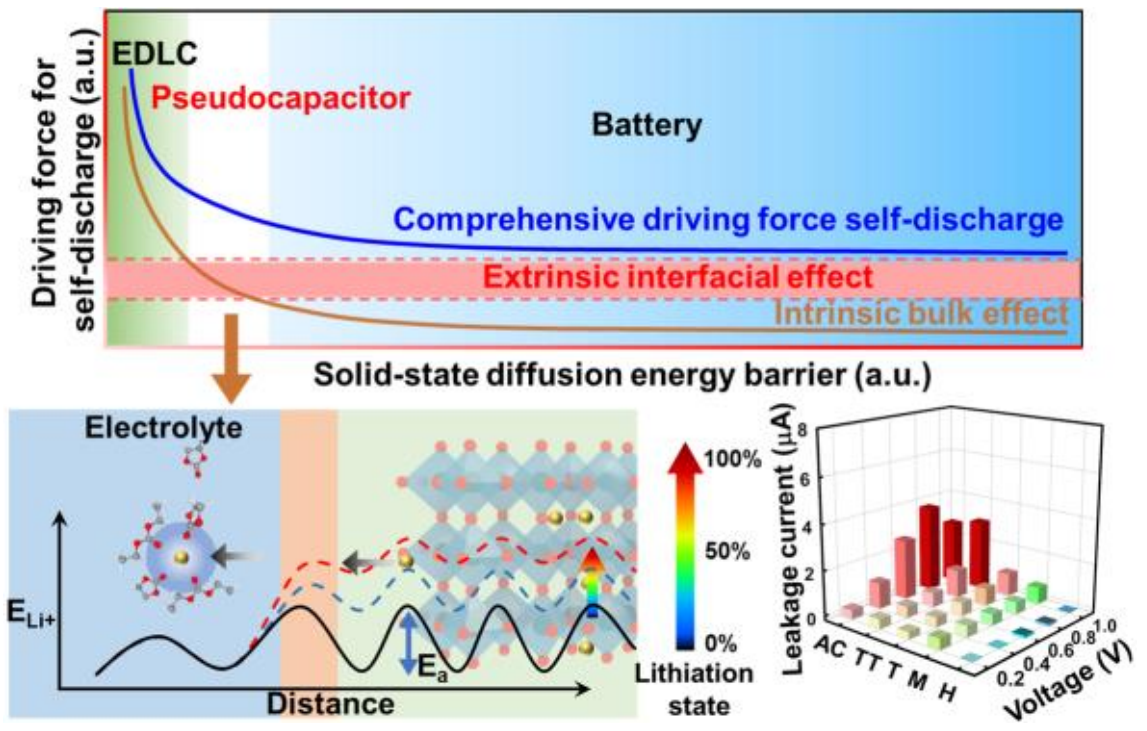

Self-discharge is a significant challenge faced by many high-rate energy storage devices.[1-4] While self-discharge is slow and therefore less discussed for lithium-ion batteries (4% per month), it has been identified in the nickel-cadmium batteries (20% per month) and nickel-metal hydride batteries (25% per month). Noteworthy, the self-discharge rate of supercapacitors reaches 1.8% per day, which severely limits its application.[5, 6] Current understanding has attributed the mechanisms of self-discharge for supercapacitors to three processes: faradaic reactions, charge redistribution, and ohmic leakage.[7, 8] The faradic reactions are caused by the presence of impurities in the electrolyte that are capable of redox reactions.[9, 10] Charge redistribution refers to the transfer loss of absorbed charged ions due to the concentration gradient.[11] Ohmic leakage is related to the incomplete isolation of the positive and negative electrodes.[12, 13] Accordingly, numerous studies have demonstrated the modification of individual components (mostly electrode or electrolyte) for suppressing self-discharge.[14-17] However, even with ingenious manufacturing technology for minimizing possible faradaic reactions, charge redistribution, and ohmic leakage, the self-discharge process of supercapacitors is still much more severe than that of lithium-ion batteries. Theoretically, the three mechanisms mentioned above are general and apply equally to all energy storage devices. It's rational to hypothesize that other intrinsic mechanisms should exist specifically for energy storage devices with significant self-discharge, like supercapacitors. Unfortunately, further clarification is complex because the two electrode/electrolyte interfaces (at cathode and anode, respectively) play an essential role for self-discharge while they involve different and complex interfacial processes, thus being hard to characterize and interpret.

Fundamentally, the essence of self-discharge is the loss of energy which is accompanied by the departure of the charge carrier from the electrode to the electrolyte driven by the Gibbs free energy.[18, 19] Therefore, the diffusion process for charge carriers is intuitively playing a decisive role in self-discharge behavior.[20] To clarify this effect, it's preferable to establish a capacitive energy storage device which involves only one type of charge carrier to monitor the self-discharge process. This is challenging in conventional carbon-based electrochemical double-layer capacitors (EDLCs) because the cathode and anode work at different voltages and adsorb different ions. Recently, the emergence of intercalation pseudocapacitive materials provides a series of unconventional working mechanisms which enable the construction of supercapacitors with innovative configurations. As a typical example, lithium-ion can insert into the niobium oxide bulk which is similar to a battery process, while its unique continuous phase-transition process provides a sloped charge/discharge curve which is similar to a supercapacitor.[21, 22] Therefore, we have conceived a "conjugately configured supercapacitor" based on pairs of pre-lithiated niobium oxide forming a $Li_xNb_2O_5$ vs. $Li_{2-x}Nb_2O_5$ structure. Conjugation is a concept widely used in different disciplines. It describes a matched pair of subjects that can interconvert under specific rules.[23-25] Considering the charge/discharge process of the conjugated supercapacitor involves the reversible transfer of a single type of charge carrier, it provides an ideal platform to study the self-discharge mechanism.

Herein, we prepared $Nb_2O_5$ with various phases, including TT-, T-, M- and H-$Nb_2O_5$. The solid-state diffusion process of the charge carrier was elucidated by *ab initio* calculations. By assembling pairs of pre-lithiated materials to construct conjugated supercapacitors, the self-discharge behavior of these devices was compared and associated with the solid-state diffusion energy barrier of the charge carrier. Theoretical and experimental analysis indicate that the solid-state diffusion energy barrier dominates for EDLCs and pseudocapacitors, and it determines the self-discharge process with an exponential form. The conjugated supercapacitor based on H-$Li_xNb_2O_5$ vs. H-$Li_{2-x}Nb_2O_5$ maintained 59% of the open-circuit voltage after 375 h, which is two orders more stable than AC vs. AC (10%, 11 h). This work is expected to unravel the influence of the solid-state diffusion energy barrier on the self-discharge process, thus providing a general guidance for suppressing self-discharge.

Self-discharge exists universally for all energy storage devices and becomes specifically noticeable for fast charge/discharge systems like supercapacitors.[26, 27] Current understandings have mainly attributed the driving force of self-discharge to extrinsic interfacial effect (faradaic reactions, charge redistribution, and ohmic leakage). These extrinsic interfacial effects are expected to be determined by manufacturing conditions (purity of electrolyte, elimination of micro short-circuit, etc.) and apply equally to different energy storage systems (Fig. 1a).[28, 29] However, alleviating the self-discharge of supercapacitors to the degree of conventional batteries are still enormously challenging, indicating there should exist intrinsic driving force (determined by the property of the active materials) which governs the self-discharge process of supercapacitors. Considering the migration of the charge carrier at one degree of freedom driven by thermal vibration (which corresponds to ~ 0.013 eV at 300 K calculated from 1/2 $kT$), the rate of this process is determined by the solid-state diffusion energy barrier with an exponential form and would dominate for materials with comparably small diffusion energy barrier. For capacitive materials with a linear correlation between the potential and the state of charge (SOC), the loss of capacity results in the change of voltage, which can be experimentally monitored. Under the combined effect of extrinsic interfacial effect and intrinsic bulk effect, the voltage vs. time curve under open circuit condition (Fig. 1b) and the leakage current curve under given applied voltage (Fig. 1c) can theoretically be established (methods given in Fig.S1). The results predict that the intrinsic bulk effect is negligible for batteries with a high solid-state diffusion energy barrier, while the extrinsic interfacial effect dominates. On the contrary, the solid-state diffusion energy barrier would significantly influence (because of its exponential form) the self-discharge process for capacitive materials. To testify the proposed mechanism, an electrochemical energy storage device with capacitive behavior (sloped charge/discharge curve) and functions with symmetric reaction environment at both electrodes (involving a single type of charge carrier) is preferred. Niobium oxides, with their unique intercalation pseudocapacitive properties, could be used to construct conjugately configured supercapacitors for observation. We first prepared niobium oxides with different phases and simulated the migration process of $Li^+$ within the materials. The primary step for the transport of $Li^+$ involves the migration from one bridging O site to another because of the following considerations:[30]

a) $Li^+$ can migrate directly between neighboring bridge sites.

b) Sufficient void size for Li-ion movement.

c) No considerable electrostatic barrier or resistance from Nb or O atoms.

Consequently, it can be utilized as a helpful indicator to plot the outline of the complete diffusion path for $Li^+$ in various $Nb_2O_5$ lattices (Fig. S2). The energy changes induced by Li-ion migration at various diffusion sites were quantified for different phases (Fig. 2a-d, Fig. S3 and Table S1). The rate-determining step (RDS) can then be identified and marked with red lines in Fig. 2a-d. The RDS can determine the general diffusion coefficient within the continuous transport chain (which is analogous to a reaction chain). Thermodynamically, the diffusion of $Li^+$ in TT-$Nb_2O_5$ has the lowest energy uphill in the RDS, i.e., the highest diffusion probability. In contrast, the $Li^+$ are more challenging to migrate in the H-$Nb_2O_5$ and M-$Nb_2O_5$ lattice.

To further quantify the kinetics, we have calculated the diffusion energy barrier of the RDS for various lattices (see the diffusion pathway animated in Movie S1). The diffusion of $Li^+$ also has the minimum barrier of 0.5 eV in TT-$Nb_2O_5$ and the highest barrier of 0.85 eV in H-$Nb_2O_5$. Noteworthy, transition-state theory describes the $Li^+$ migration based on diffusion barriers in the ground state (0 K). To provide further insights into the diffusion of $Li^+$ at room temperature (300 K), ab initio molecular dynamics (AIMD) was carried out. Given with constant temperature and volume, the Einstein relation can be used to effectively calculate the diffusivity from AIMD simulations for the canonical ensemble (NVT). After computing the mean square displacement (MSD), the slope of the MSD *vs.* simulation time is $2dD$, where $d$ is the is the dimensionality of the diffusion and $D$ is the diffusion coefficient. At the same time scale, the calculated MSD curves (Fig. S4) show that Li-ions in TT-$Nb_2O_5$ have the fastest diffusion coefficient at a finite temperature of 300 K. The vibrant microscopic Li-ions diffusion mechanism for various lattices can be further visualized by AIMD (Movie S2).

To experimentally identify how the solid-state diffusion energy barrier would determine the self-discharge process, we prepared niobium oxide with similar morphologies (Figure S6) but different phases (Figure S5) by annealing the nano precursor at various temperatures.[31] Afterwards, coin cells were assembled using lithium metal foil as the counter electrode (Fig 3b). The EDLC-type activated carbon (YP-80F) was used as the control group for comparison. The CV, EIS and Galvanostatic charge/discharge curves (Fig. S7 and S8) of the niobium oxides show characteristic intercalation-type pseudocapacitive behaviors, thus enabling the materials to serve as ideal samples for unraveling the effect of the solid-state diffusion process on the self-discharge property.[32] Compared to straight sloped charge/discharge curve (capacitive) for M-$Nb_2O_5$, T-$Nb_2O_5$ and TT-$Nb_2O_5$ phase, the H-$Nb_2O_5$ phase annealed at higher temperature (1100°C) shows partially plateaued charge/discharge curve (battery-like) and has the highest diffusion energy barrier, which is consistent with the DFT calculation.[33] To evaluate the self-discharge property, the cells were stabilized for 10 charge/discharge cycles and then kept at 3.0 V or 1.0 V for 2 h to reach the lithium-empty or lithium-full state at the initial stage. Then we measured the change of open circuit potential (OCP) during aging (Fig 3a). The voltage of the half cells instantly changed from 3.0 V to about 2.3 V or from 1.0 V to about 1.2 V. Afterward, the voltage became relatively stable even when aged for 1000 h. To further analyze the stability of the materials, the leakage currents were collected by recording the currents at various times when a constant voltage of 3 V was applied (Fig 3c and 3d).[34] The results indicate that the current became stable after about 10 hours. The activated carbon based on the EDLC mechanism shows significantly higher leakage current (about 4.0 μA) than that of the niobium oxides based on the intercalation pseudocapacitive process. Among all niobium oxides, the H-$Nb_2O_5$ shows the lowest leakage current of about 1.0 μA, and the T-$Nb_2O_5$ shows the highest leakage current of about 2.2 μA. Besides, the leakage current decreases with lower applied voltage (Fig S9).[35] The above results indicate that for the materials assembled using lithium metal as the counter electrode, a higher solid-state diffusion barrier would result in lower self-discharge for such devices.

The experimental understanding of the correlation between the solid-state diffusion barrier of charge carrier and the self-discharge process of the electrochemical energy storage device has been hindered when using conventional EDLC supercapacitor or battery systems as the studying group because the EDLC supercapacitors are based on interface process while conventional batteries essentially involve two different materials to serve as the cathode and anode (using the same material as cathode and anode cannot generate voltage and storage energy).[36] The recent emergence of intercalation pseudocapacitive materials represented by niobium oxide could experience a unique continuous phase-transition process during lithium insertion/extraction,[37] thus providing a sloped charge/discharge curve and enabling the establishment of a conjugately configured device that could serve as an ideal candidate to shed light on how the solid-state diffusion of charge carrier determines the self-discharge process.[38-40] Through the pre-lithiation method, pairs of pre-lithiated niobium oxide can be coupled to store energy where the same charge carrier shuttles between the electrodes (Fig S10 and S11). This device involves only one type of charge carrier and similar electrode/electrolyte interface on both sides, which minimizes the variables for studying the influence of diffusion barrier. We then measured the change of OCP to evaluate the stability of the conjugated supercapacitor (Fig 4a). The supercapacitors were stabilized for 10 charge/discharge cycles, kept at 1.0 V for 2 h, and then aged for testing. The voltage of the conjugated device based on H-$Nb_2O_5$, M-$Nb_2O_5$, T-$Nb_2O_5$ and TT-$Nb_2O_5$ decayed to 0.59, 0.37,0.03, and 0.16 V during an aging time of 375 h, which is consistent of the solid-state diffusion energy barrier as predicted by DFT. For comparison, the voltage of EDLC type AC vs. AC with low energy barrier for charge carrier quickly decayed from 1.0 V to 0.0 V in only 50 h. We further carried out a leakage current test that can reflect the rate of self-discharge when the device is applied with a constant voltage of 1.0 V (Fig 4c and 4d).[41] The leakage current (the recorded current after aging the device at the given voltages for more than 20 h) of the conjugated devices is positively related to the applied voltage (Fig. S12). When the applied voltage was kept at 1.0 V, the leakage current of the conjugated device based on activated carbon (~ 4.0 μA) is higher than other devices based on niobium oxides. The H-$Li_xNb_2O_5$ vs. H-$Li_{2-x}Nb_2O_5$ device shows a negligible leakage current (recorded as about 0.0 μA under the resolution of our LAND electrochemical test system with the upper measuring range of 5.0 mA). To directly visualize the difference in the self-discharge performance, we charged the conjugated supercapacitor based on the H-phase or the TT-phase. They could both light the red LED with similar brightness at the initial stage. On the contrary, the devices aged at open-circuit conditions for various times show significantly different performance. After 60 h, the H-$Li_xNb_2O_5$ vs. H-$Li_{2-x}Nb_2O_5$ conjugated supercapacitor could still light the red LED while the TT-$Li_xNb_2O_5$ vs. TT-$Li_{2-x}Nb_2O_5$ device failed (Fig 4e and 4f). These results provide solid evidence for the deterministic role of solid-state diffusion energy barrier of charge carrier on the self-discharge performance of energy storage systems.

By coupling pairs of pre-lithiated niobium oxides with similar intercalation pseudocapacitive properties but different phases, we have constructed conjugatedly configured supercapacitor as the platform to unravel the effect of the solid-state diffusion energy barrier of charge carriers on the self-discharge process. The theoretical analysis combined with experimental verification indicates that the self-discharge is not only driven by the extrinsic interfacial effect (faradaic reactions, charge redistribution, and ohmic leakage) as previously reported but also by the intrinsic solid-state diffusion energy barrier. This understanding is crucial for designing electrochemical energy storage devices with suppressed self-discharge. We demonstrated the conjugated $Li_xNb_2O_5$ vs. $Li_{2-x}Nb_2O_5$ shows significantly higher stability than AC vs. AC symmetric supercapacitors. In particular, using H-$Li_xNb_2O_5$ with the highest solid-state energy barrier as the active materials is efficient for further enhancing the resistance to self-discharge. This work is expected to provide general guidance for suppressing self-discharge for fast charge/discharge electrochemical energy storage devices.

## ■ Experimental section

*Synthesis of the $Nb_2O_5$ raw materials.* The uniform precursor of $Nb_2O_5$ was synthesized by the solvothermal-assisted sol-gel method. Firstly, 2 mmol $NbCl_5$ (Energy Chemical) was dissolved into 40 mL ethanol. The solution gradually turned into a colorless sol during the 15 min stirring process. Then the solution was stirred for another 1 hour after adding 10 mL water and turned colorless to milky due to the hydrolyzation of niobium ions. Finally, the opaque milky solution was transferred into a Teflon-lined autoclave and sealed in a steel container. A cylindrical monolithic milky gel formed after heating at 180°C for 12 h. The obtained gel was washed with deionized and ethanol three times repeatedly, separated by centrifugation, and dried at 80°C. The amorphous precursor powder was obtained by grinding the dried xerogel. Then the precursor powders were annealed at 500, 600, 850, and 1100°C for 1 h in the air, correspondingly to obtain the TT-$Nb_2O_5$, T-$Nb_2O_5$, M-$Nb_2O_5$ and H-$Nb_2O_5$.

*Materials characterization and electrochemical tests.* Structural characterizations: Phase and morphology were examined by X-ray diffraction (XRD, Bruker D8 Advance), scanning electron microscope (SEM, FEI/INSPECT F50), and transmission electron microscope (TEM, Tecnai G2 F20 at 200 kV).

EC tests: The cells were aged for 24 h before the test. The cycles and self-discharge were tested on the LAND CT3001A battery tester. The cyclic voltammetry (CV) and electrochemical impedance spectroscopy (EIS) were tested with the CHI660E electrochemical workstation (CH Instruments Ins).

*Assembly of Symmetric Supercapacitor based on pre-lithiated $Nb_2O_5$.* Electrodes preparation: The working electrodes were obtained by coating the slurry on the Cu foil. The slurry consisted of 80 wt% activated materials, 10 wt% acetylene black (AB, Canrd) and 10 wt% poly (vinylidene fluoride) (PVDF, Canrd) dissolved in N-methyl pyrrolidinone (NMP, Canrd). The electrodes were vacuum dried at 110°C after being blast dried at 80°C. Finally, the electrodes were punched into round disks with 12 mm diameter.

2032 coin cells were assembled in a glove box filled with pure argon gas. Celgard 2500 microporous membrane was used as the separator and the electrolyte was 1.0 M $LiPF_6$ in EC: DMC (3:7). For pre-lithiation, lithium foil was used as the counter electrode ($Nb_2O_5$ vs. $Li/Li^+$) to assemble half cells. The half cells were first stabilized by running for 10 cycles at 1.0-3.0 V. Then electrodes at lithiated state (discharged to 1.0 V) or delithiated state (charged to 3.0 V) were obtained by disassembling the half cells. The delithiated and lithiated $Nb_2O_5$ electrodes were used as cathode and anode to assemble the symmetric device. As a control group, EDLC AC vs. AC coin cells were assembled. To visualize the self-discharge performance, three coin cells connected in series were used to light LEDs after aging for the given time.

*Simulation methods.* We used the Vienna ab initio simulation software (VASP)[42] to conduct ab initio calculations based on density functional theory (DFT) and the projector augmented wave (PAW) formalism[43, 44]. The Perdew-Burke-Ernzerhof (PBE) functional[45] was used for the exchange-correlation function. The van der Waals (vdW) corrections were considered by DFT-D3.[46] Geometry optimizations were carried out by employing a 450 eV kinetic energy cutoff and a force convergency of less than 0.05 eV. The transition states were determined using the climbing-image nudged elastic band (Cl-NEB) technique.[47] The conventional lattices of H-$Nb_2O_5$, M-$Nb_2O_5$, T-$Nb_2O_5$, and TT-$Nb_2O_5$ (see supplementary CIF files) were constructed with the equivalent Monkhorst-Pack k-points density of 0.04 1/Å, corresponding to 5 x 1 x 1, 1 x 1 x 6, 4 x 1 x 6 and 7 x 2 x 2. Furthermore, considering the expenses, ab initio molecular dynamics (AIMD) simulations of 2 ps with a time step of 2 fs at the canonical ensemble (NVT) system of 300 K were able to analyze the diffusion coefficient differences between various structures. All the structures or models were visualized using the VESTA software[48] or the OVITO program[49].

## ■ ASSOCIATED CONTENT

**Supporting Information**

This information is available free of charge online.

Supplementary notes, the results of materials characterization and electrochemical tests; the model and process of simulation. (docx.)

Supporting video 1, the diffusion path of rate-determining step for $Li^+$ in niobium oxides. (mp4.)

Supporting video 2, the motion of vibrant microscopic $Li^+$ in niobium oxides. (mp4.)

## ■ AUTHOR INFORMATION

**Corresponding Author**

**Xiang Ge** – Department of Materials and Metallurgy, Guizhou University, Guiyang, Guizhou 550025, China; E-mail: xge@gzu.edu.cn

**Authors**

**Xiaohui Yan** – Department of Materials and Metallurgy, Guizhou University, Guiyang, Guizhou 550025, China

**Yue He** – Department of Chemical & Materials Engineering, The University of Auckland, Auckland 1142, New Zealand

**Xuncheng Liu** – Department of Materials and Metallurgy, Guizhou University, Guiyang, Guizhou 550025, China

**Siqi Jing** – Department of Materials and Metallurgy, Guizhou University, Guiyang, Guizhou 550025, China

**Jiajian Guan** – Department of Chemical & Materials Engineering, The University of Auckland, Auckland 1142, New Zealand

**Wei Gao** – Department of Chemical & Materials Engineering, The University of Auckland, Auckland 1142, New Zealand

**Sudip Ray** – Department of Chemical & Materials Engineering, The University of Auckland, Auckland 1142, New Zealand; New Zealand Institute for Minerals to Materials Research, Greymouth 7805, New Zealand

**Yige Xiong** – Department of Materials and Metallurgy, Guizhou University, Guiyang, Guizhou 550025, China

**Taibai Li** – Department of Materials and Metallurgy, Guizhou University, Guiyang, Guizhou 550025, China

**Author Contributions**

$^{+}$ Xiaohui Yan and $^{+}$ Yue He contributed equally to this work.

**Note**

The authors declare no conflict of interest.

**Acknowledgments**

This work was supported by the National Natural Science Foundation of China (52262030) and Natural Science Foundation of Guizhou Science and Technology Department (QKHJC-ZK[2021]-YB257).

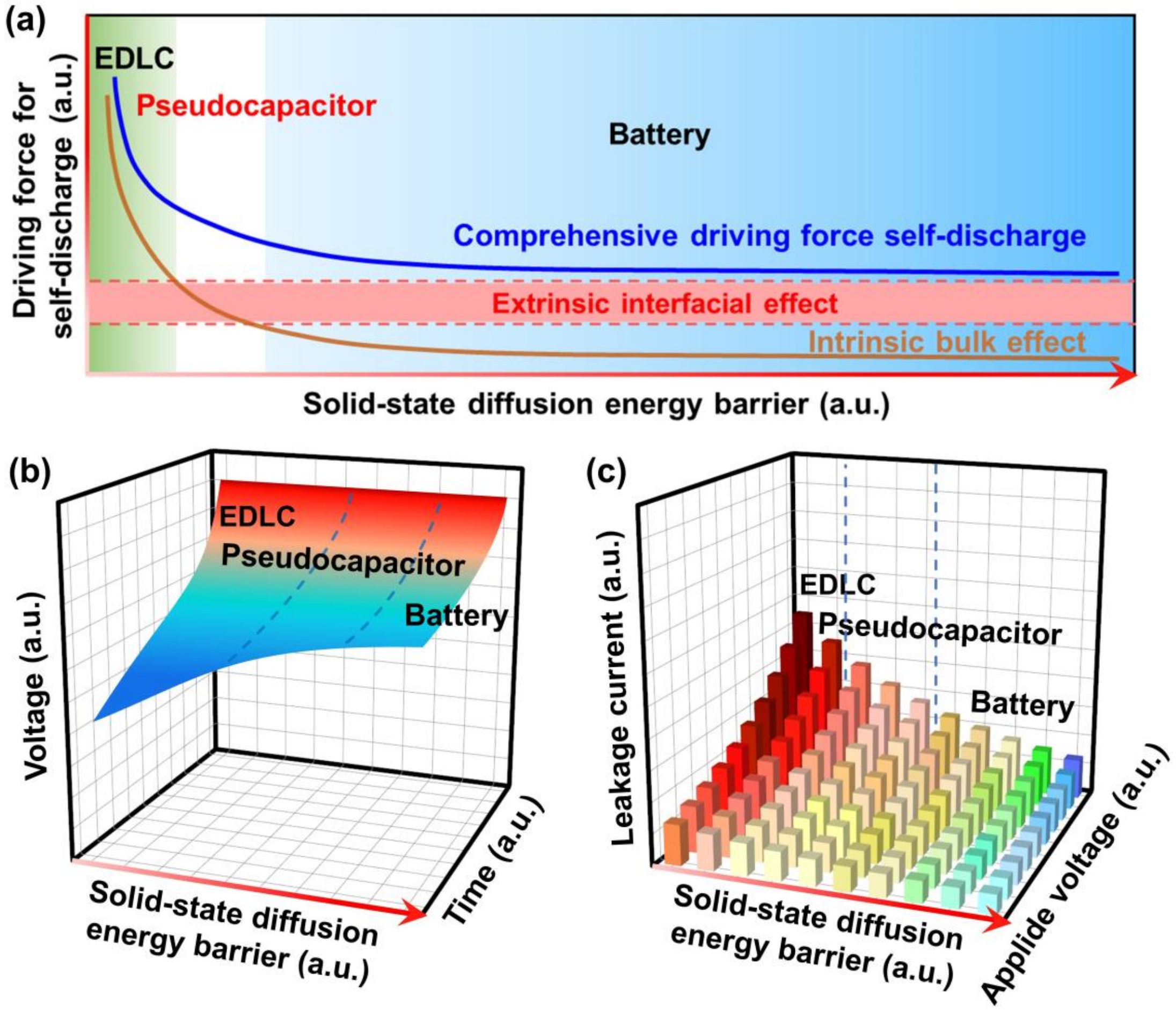

**Figure 1.** Schematic illustration showing the theoretical influence of the solid-state diffusion energy barrier on self-discharge. (a) The comprehensive driving force for self-discharge is determined by the intrinsic bulk effect and extrinsic interfacial effects, where the bulk intrinsic effect would dominate for materials with lower diffusion energy barrier. (b) shows the theoretical voltage-time curve for materials with different diffusion energy barriers. (c) gives the theoretical leakage current at various applied voltages for materials with different diffusion energy barriers.

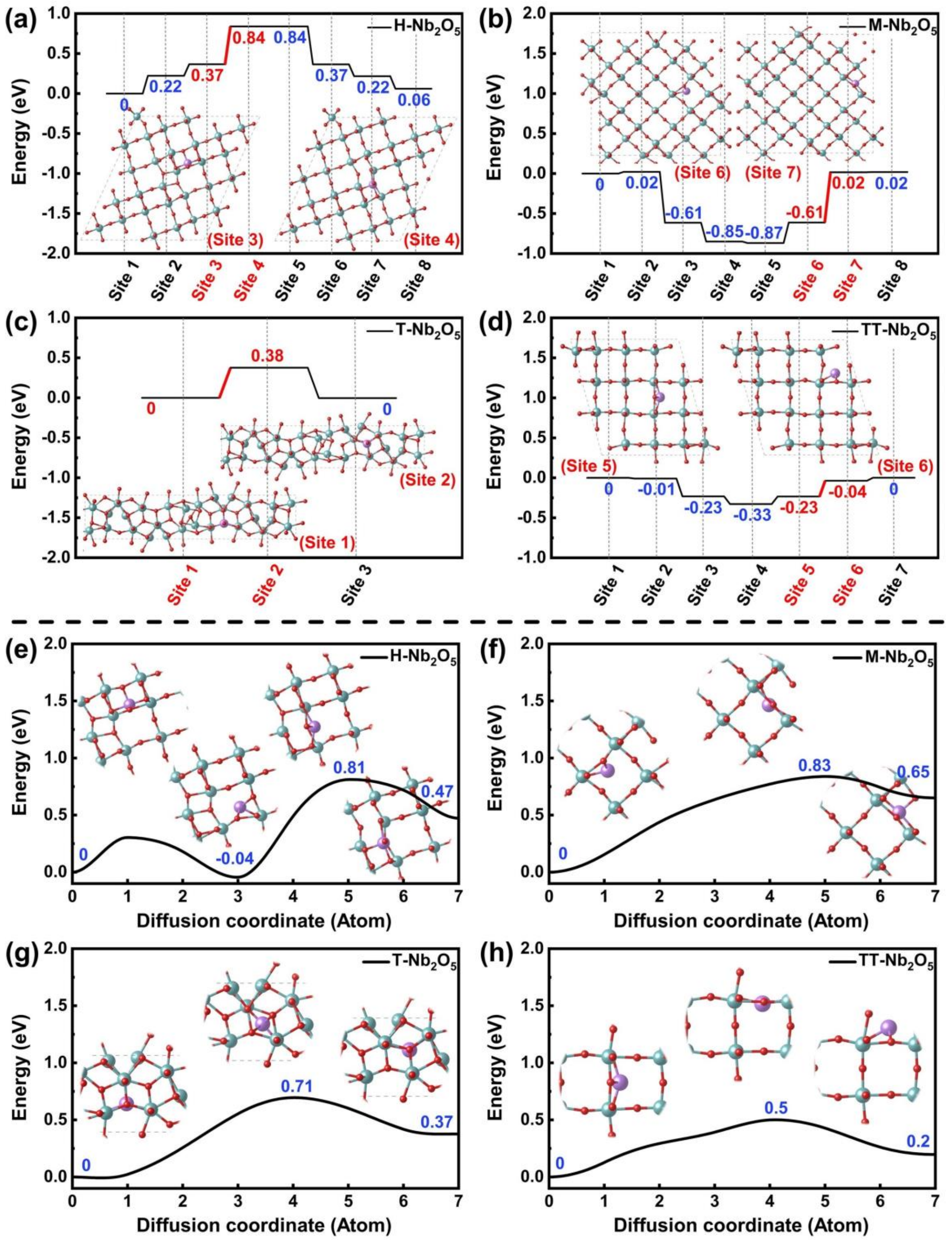

**Figure 2.** Energy landscape for the diffusion of $Li^+$ in different niobium oxides based on DFT calculations. (a-d) show the energy changes at various diffusion sites in $H\text{-}Nb_2O_5$, $M\text{-}Nb_2O_5$, $T\text{-}Nb_2O_5$, and $TT\text{-}Nb_2O_5$ from the thermodynamic view; (e-h) show the diffusion barrier of the rate-determining step in $H\text{-}Nb_2O_5$, $M\text{-}Nb_2O_5$, $T\text{-}Nb_2O_5$ and $TT\text{-}Nb_2O_5$ from the kinetic aspect.

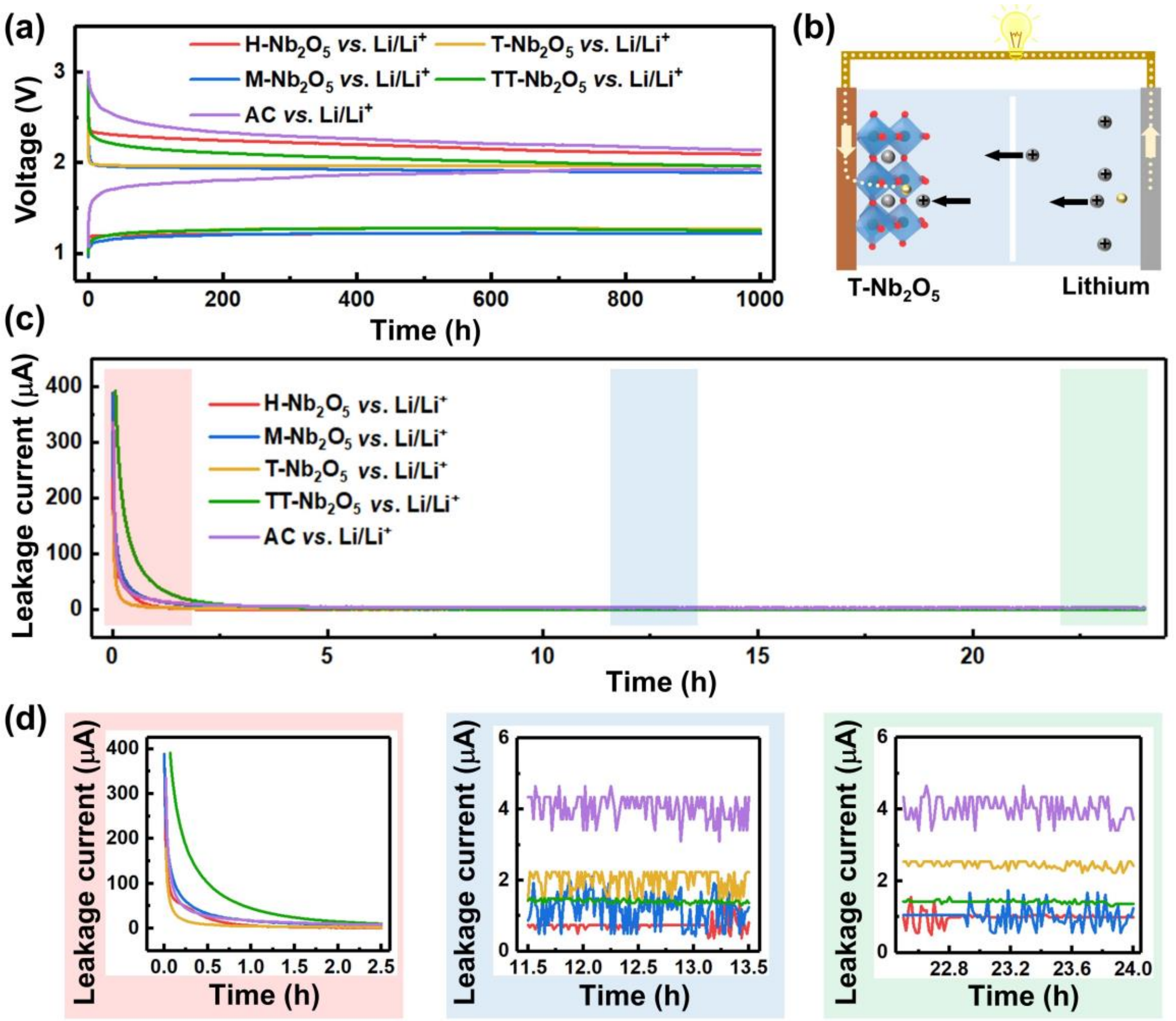

**Figure 3.** The self-discharge performance of the materials tested using lithium metal as the counter electrodes. (a) compares the self-discharge performances of H-$Nb_2O_5$, M-$Nb_2O_5$, T-$Nb_2O_5$, TT-$Nb_2O_5$ and activated carbon at initially delithiated or lithiated state. (b) shows the configuration of the half cells. (c and d) compare the leakage current curves of H-$Nb_2O_5$, M-$Nb_2O_5$, T-$Nb_2O_5$, TT-$Nb_2O_5$ or activated carbon half cells. (d) give the enlarged view at different stages of (c).

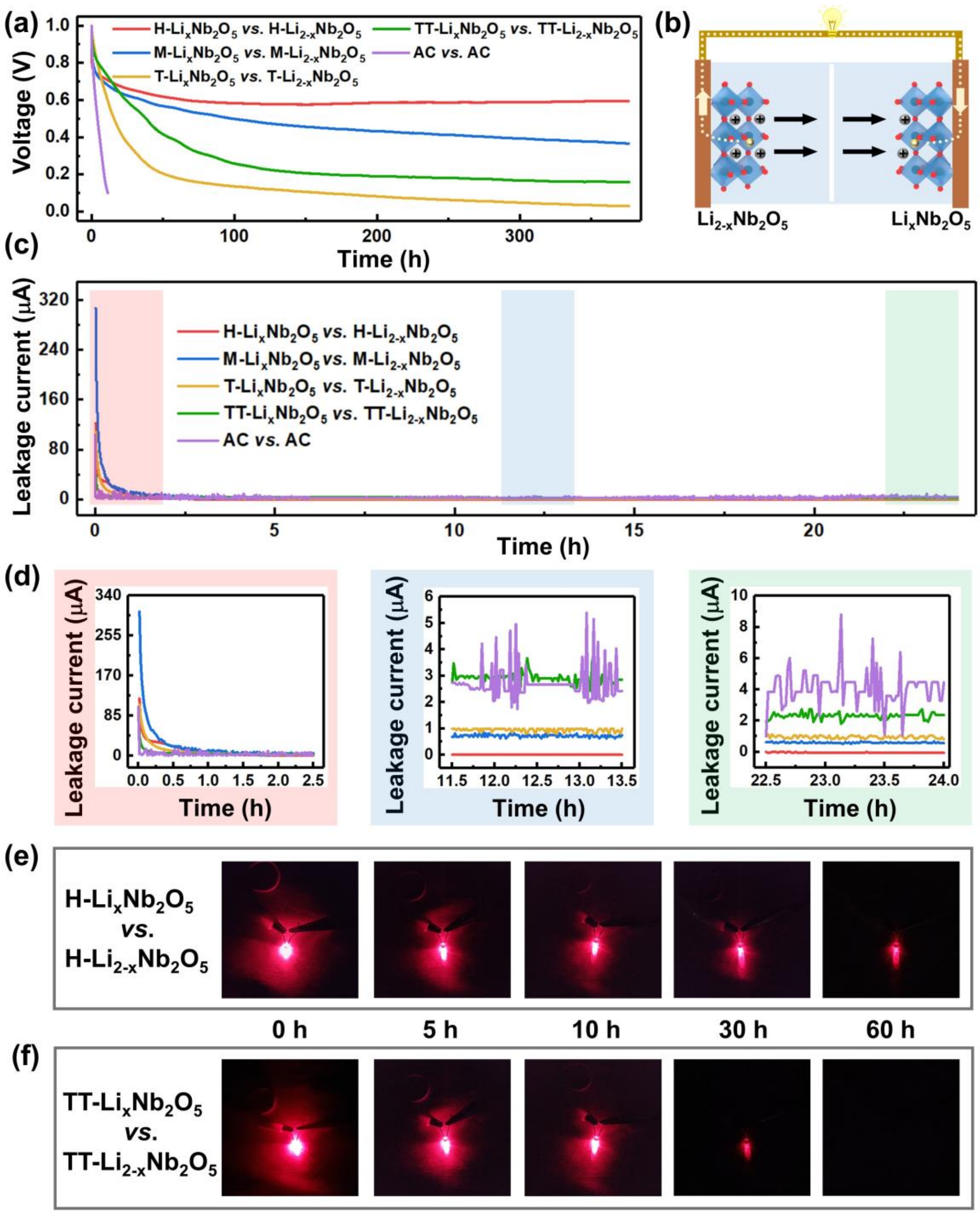

**Figure 4.** The self-discharge and electrochemical performance of symmetric supercapacitors. (a) compares the self-discharge performance of symmetric supercapacitor based on pairs of H-$Nb_2O_5$, M-$Nb_2O_5$, T-$Nb_2O_5$, TT-$Nb_2O_5$ or activated carbon. (b) shows the configuration for such symmetric supercapacitors. (c and d) compare the leakage current curves of symmetric supercapacitor based on pairs of H-$Nb_2O_5$, M-$Nb_2O_5$, T-$Nb_2O_5$, TT-$Nb_2O_5$ or activated carbon. (d) is the partial enlarged drawing of (c). (e and f) show the red LEDs lighted by symmetric supercapacitor of H-$Nb_2O_5$ (e) and TT-$Nb_2O_5$ (f) at different rest times.

**Supporting Information**

# Unravelling the deterministic effect of the solid-state diffusion energy barrier for charge carrier on the self-discharge of supercapacitors

*Xiaohui Yan*[+, a], *Yue He*[+, b], *Xuncheng Liu* [a], *Siqi Jing* [a], *Jiajian Guan* [b], *Wei Gao* [b], *Sudip Ray* [b, c], *Yige Xiong* [a], *Taibai Li* [a], *Xiang Ge* *

*Corresponding Author

Department of Materials and Metallurgy, Guizhou University, Guiyang, Guizhou 550025, China

E-mail: xge@gzu.edu.cn

[a] Department of Materials and Metallurgy, Guizhou University, Guiyang, Guizhou 550025, China

[b] Department of Chemical & Materials Engineering, The University of Auckland, Auckland 1142, New Zealand

[c] New Zealand Institute for Minerals to Materials Research, Greymouth 7805, New Zealand

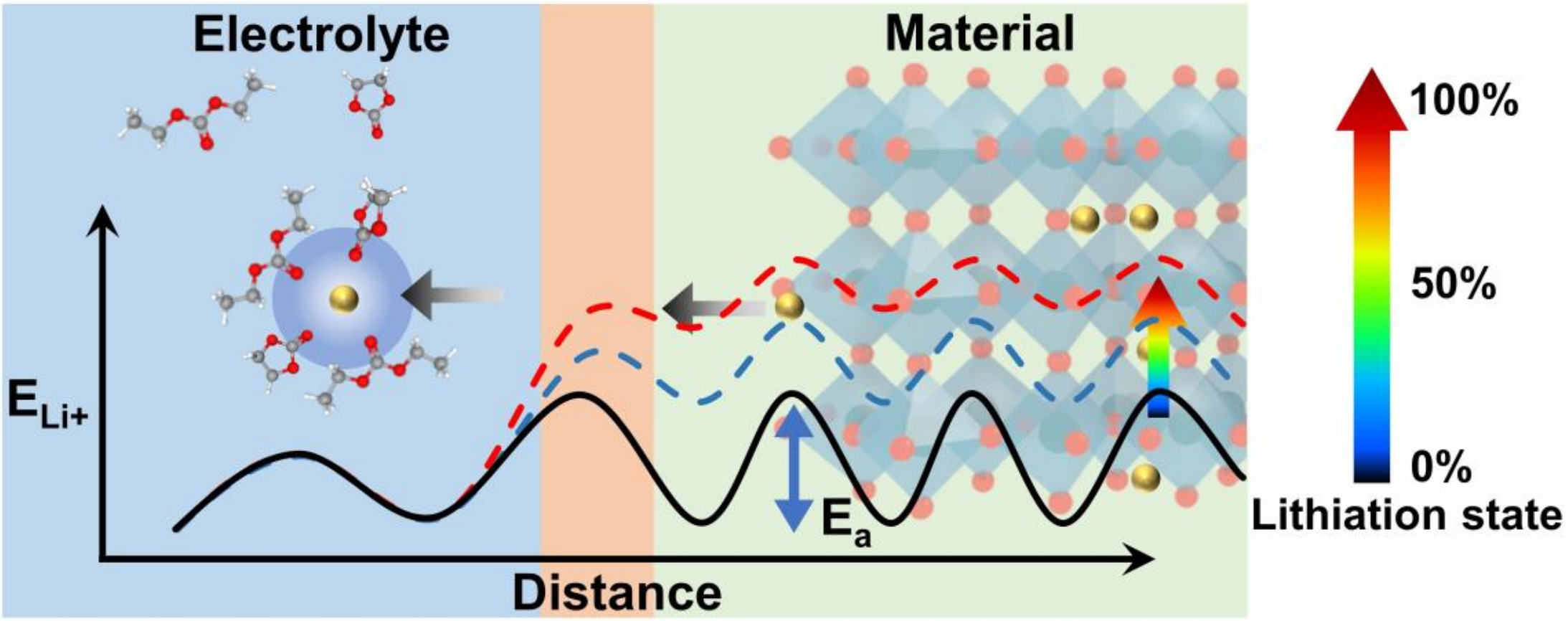

**Figure S1.** Schematic diagram showing that different lithiation state (corresponds to different voltage for pseudocapacitive materials) leads to different driving force during the migration of lithium ions during the self-discharge process.

To quantitatively unravel the effect of solid-state diffusion on the self-discharge process, the relationship between the reaction rate and the migration kinetics of the charge carrier must be established. As shown in Figure S1, considering the diffusion of a charge carrier ($Li^+$) from the lithiated bulk $Li_xNb_2O_5$ into the electrolyte during the following process:

$$Li^+ \cdot nLi_xNb_2O_5 \rightarrow Li^+(solvated\ in\ electrolyte) + nLi_xNb_2O_5 \quad [1]$$

The reaction rate ($v$) of [2] can be described using the Arrhenius equation:

$$v = kc\ exp(-\frac{E_a}{RT}) \quad [2]$$

$c$ is the concentration of the charge carrier. $E_a$ is energy barrier, and $k$ is pre-exponential factor.

The total current density causing self-discharge ($i_{total}$) include the contribution from solid-state diffusion ($i_{diffusion}$) and extrinsic factors ($i_{extrinsic}$ faradaic reactions, charge redistribution and ohmic leakage):

$$i_{total} = i_{diffusion} + i_{extrinsic} \quad [3]$$

The $i_{extrinsic}$ is mainly determined by the specific manufacturing technology of the electrochemical energy storage device. For simplification, $i_{extrinsic}$ is assumed to be constant. For $i_{diffusion}$, it can be calculated using

$$i_{diffusion} = vnFA = kcnFA \exp\left(-\frac{E_a}{RT}\right) \quad [4]$$

$n$ is the number of electrons transferred. $F$ is the Faraday's constant and $A$ is electrode area.

Therefore, the total self-discharge current density can be expressed as:

$$i_{total} = vnFA = kcnFA \exp\left(-\frac{E_a}{RT}\right) + C \quad [5]$$

$C$ is a constant and represents extrinsic interfacial effect.

During self-discharge, the energy barrier ($-E_a$) is mainly determined by the voltage

($E_{voltage}$, which is related to the lithiation state of the materials) and the diffusion energy barriers ($E_{barrier}$):

$$-E_a = f(E_{voltage}, E_{barrier}) \quad [6]$$

Therefore, the $i_{total}$ can be expressed using:

$$i_{total} = kcnFA \exp\left(\frac{f(E_{voltage}, E_{barrier})}{RT}\right) + C \quad [7]$$

Simplifying all the constants with one pre-exponential factor *m* gives:

$$I = m\, cexp\big(f(E_{voltage}, E_{barrier})\big) + C \quad [8]$$

The above equation [8] can be used to predict the leakage current at given applied voltage. To further establish the voltage vs. time curve at open circuit state during aging, we consider a capacitive material with a sloped charge/discharge curve. The change of capacity *dq* can be expressed as:

$$dq = \int I\, dt \quad [9]$$

With a nominal capacitance of *C*, the change of voltage *dv* can be expressed as:

$$dv = dq/C = \int m \exp\,(G_{driving\ force})\, dt\,/C \quad [10]$$

The voltage vs. time profile can then be theoretically plotted.

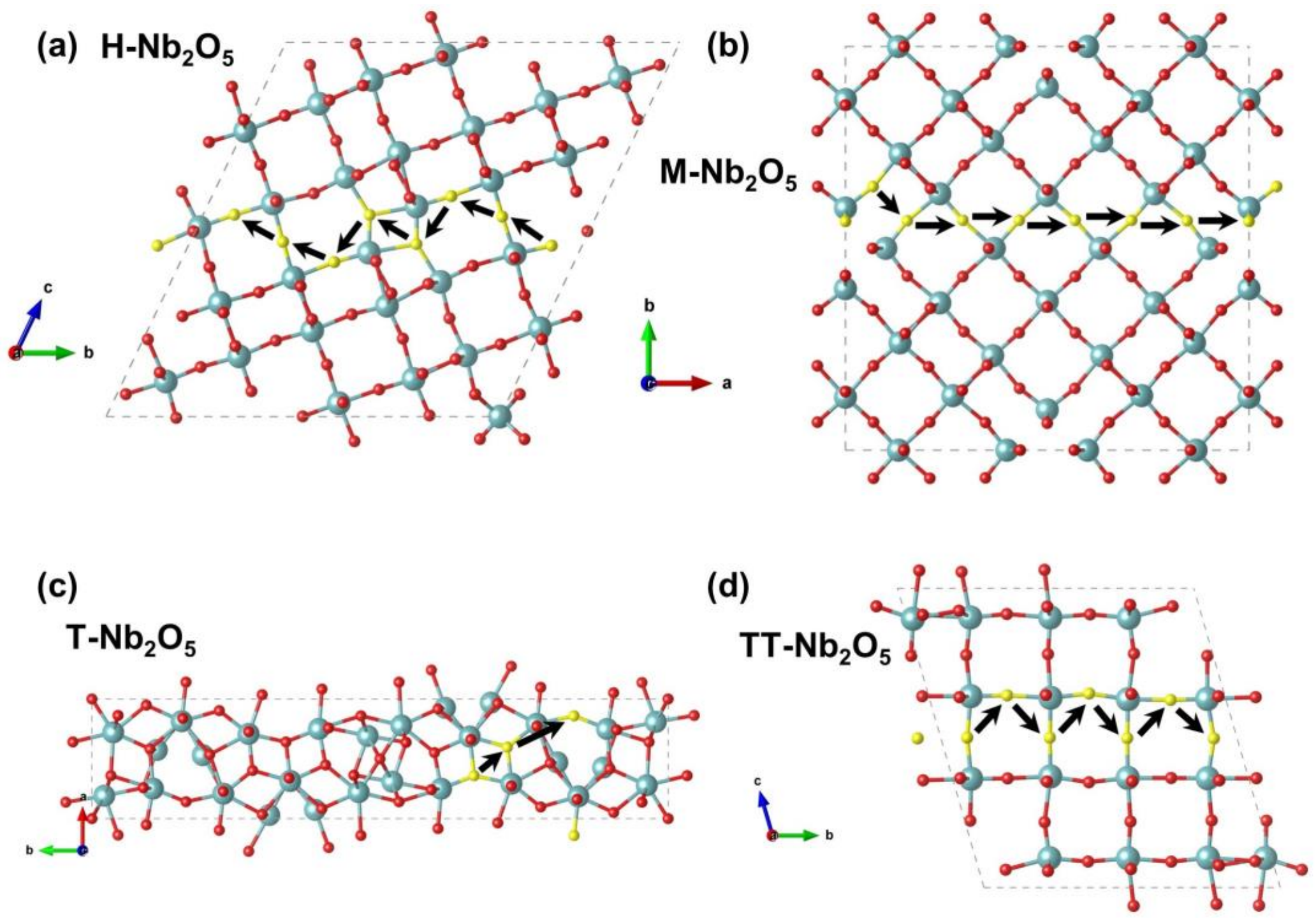

**Figure S2.** Schematic illustration showing the proposed Li-ion diffusion pathway in H-$Nb_2O_5$ (a), M-$Nb_2O_5$ (b), T-$Nb_2O_5$ (c), and TT-$Nb_2O_5$ (d). The diffusion sites at the bridge oxygen atoms are highlighted in yellow.

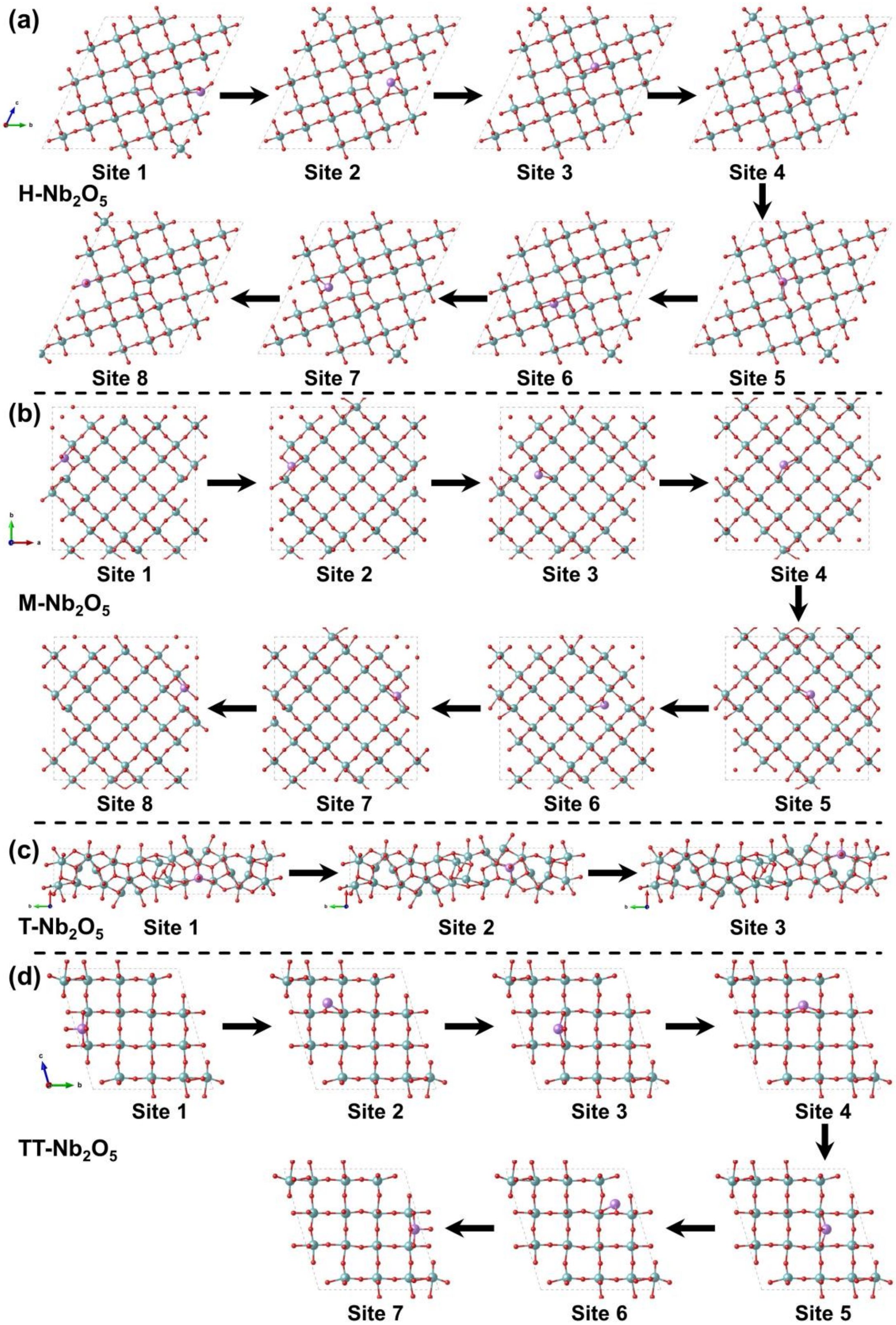

**Figure S3.** The configuration of $Li^+$ at each diffusion site in H-$Nb_2O_5$ (a), M-$Nb_2O_5$ (b), T-$Nb_2O_5$ (c), and TT-$Nb_2O_5$ (d), respectively.

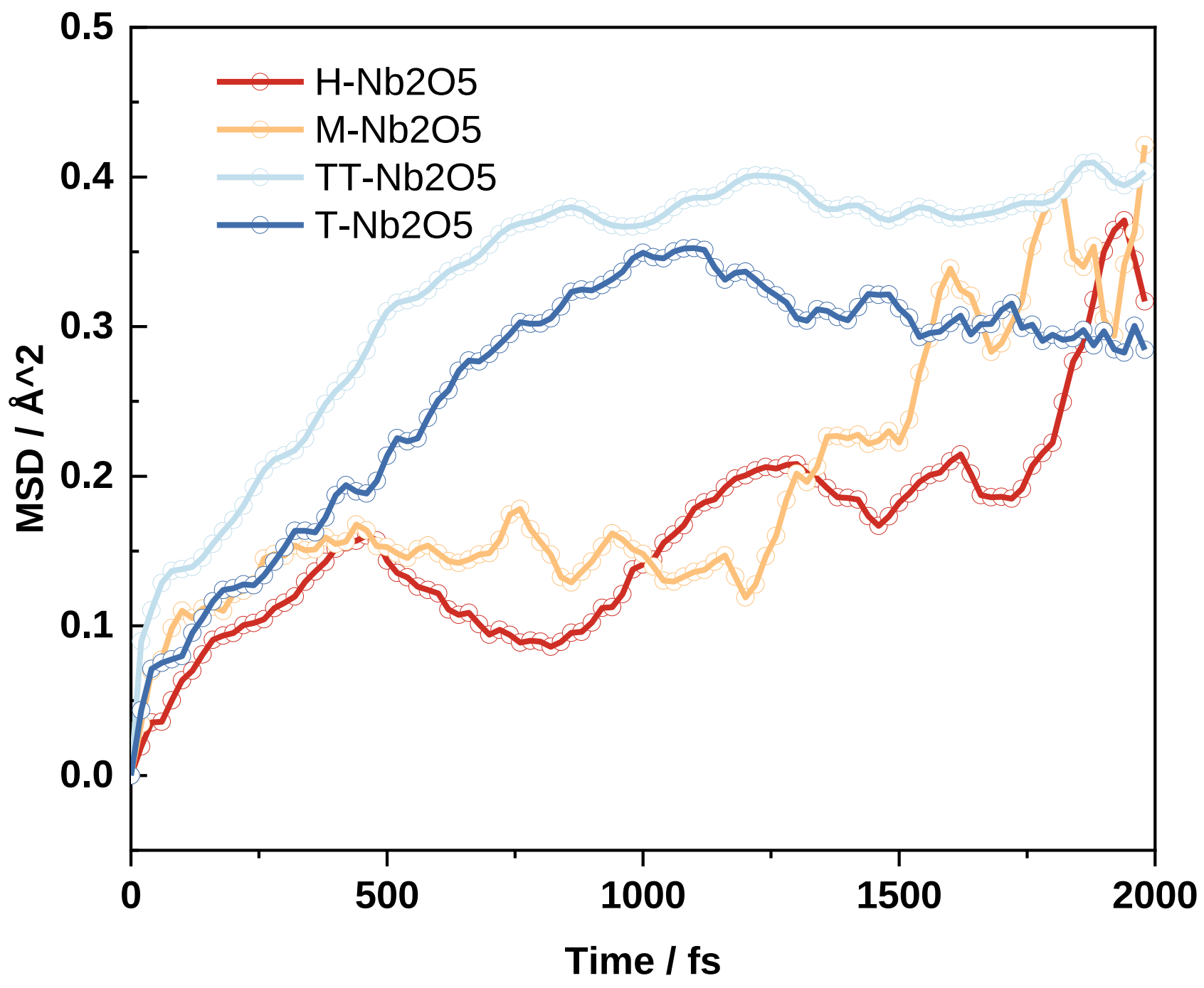

**Figure S4.** Plots of the MSD vs. the simulation time.

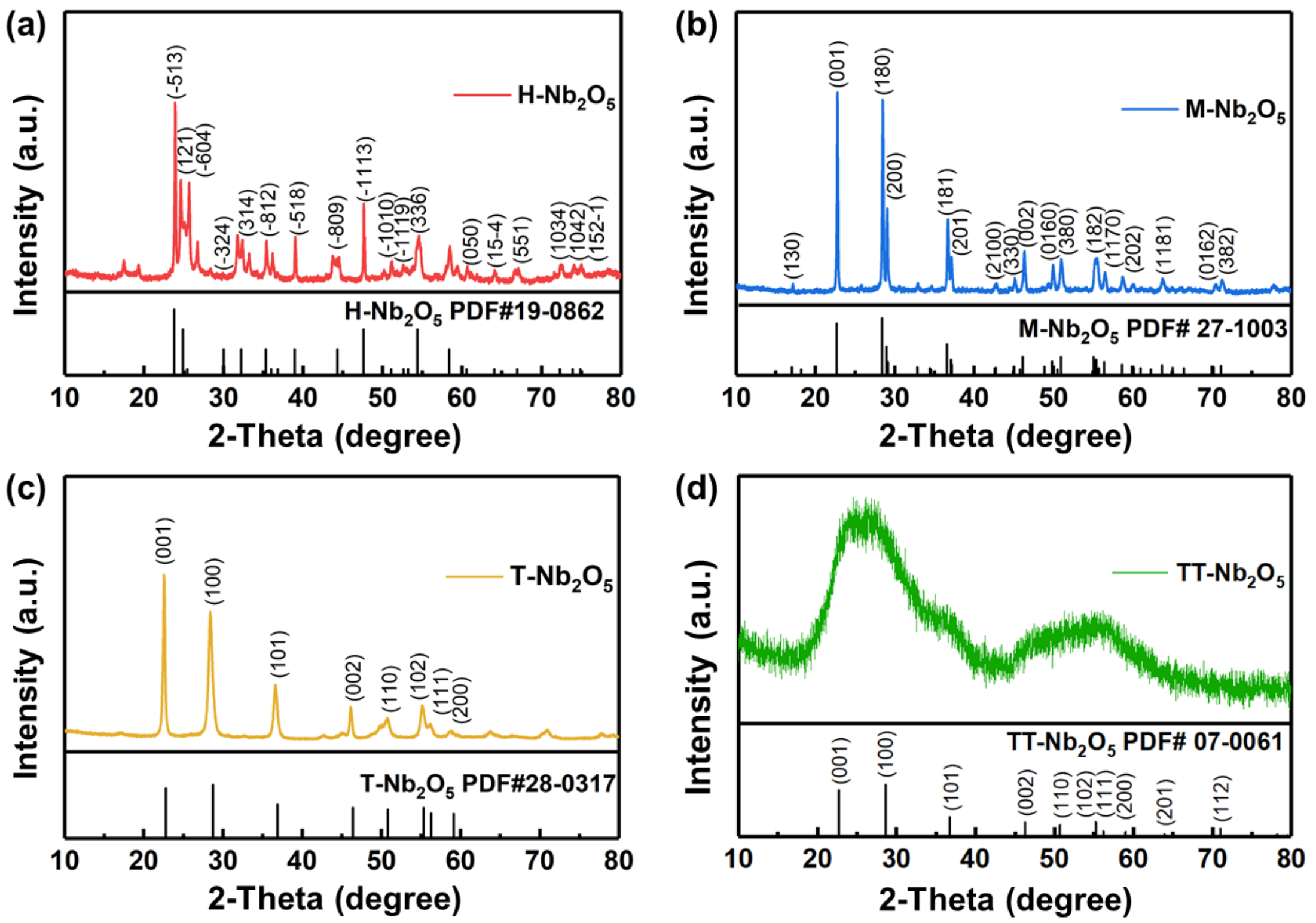

**Figure S5.** XRD patterns of the H-$Nb_2O_5$ (a), M-$Nb_2O_5$ (b), T-$Nb_2O_5$ (c) and TT-$Nb_2O_5$ (d).

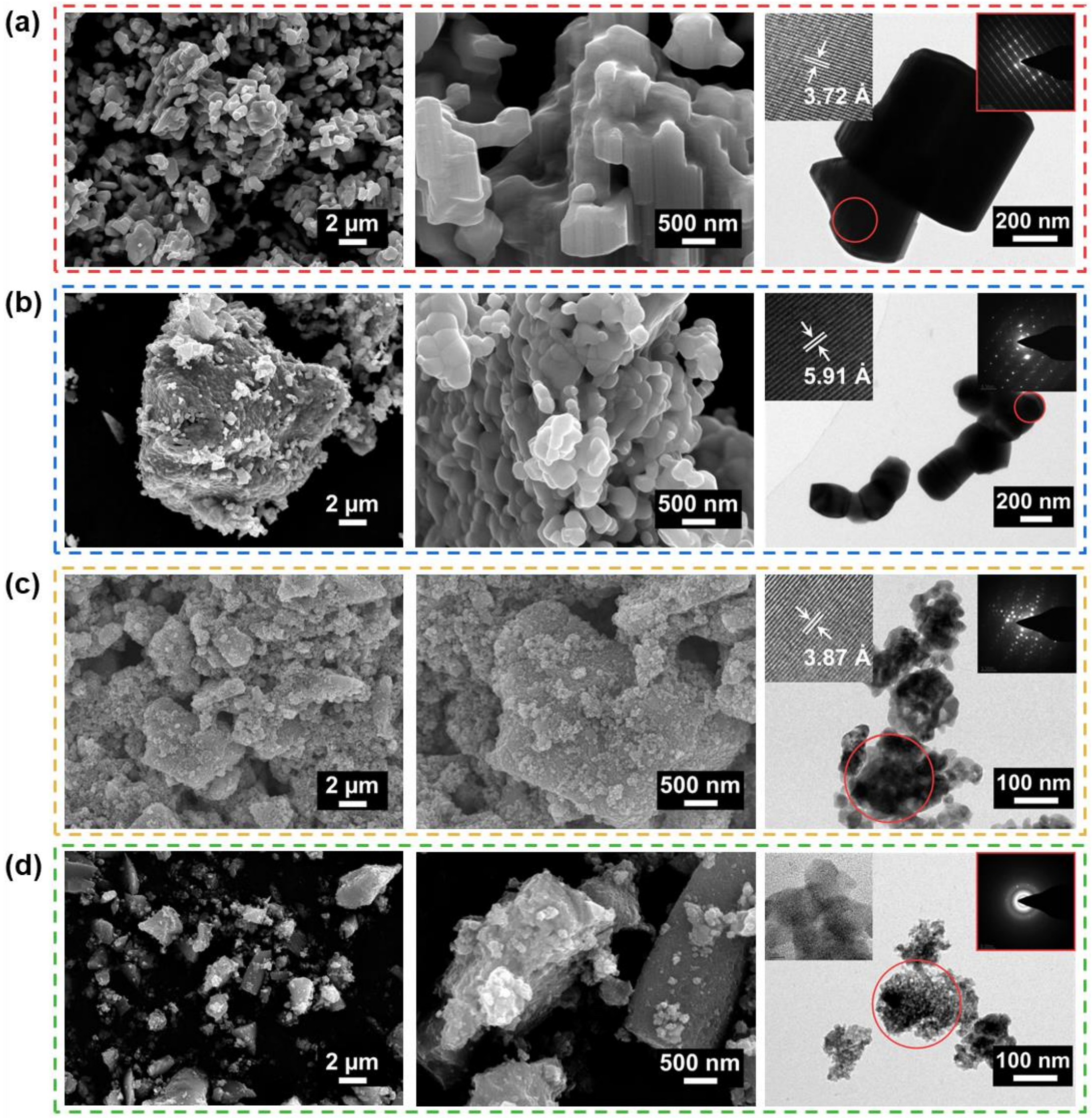

**Figure S6.** SEM and TEM images of the H-$Nb_2O_5$ (a), M-$Nb_2O_5$ (b), T-$Nb_2O_5$ (c) and TT-$Nb_2O_5$ (d).

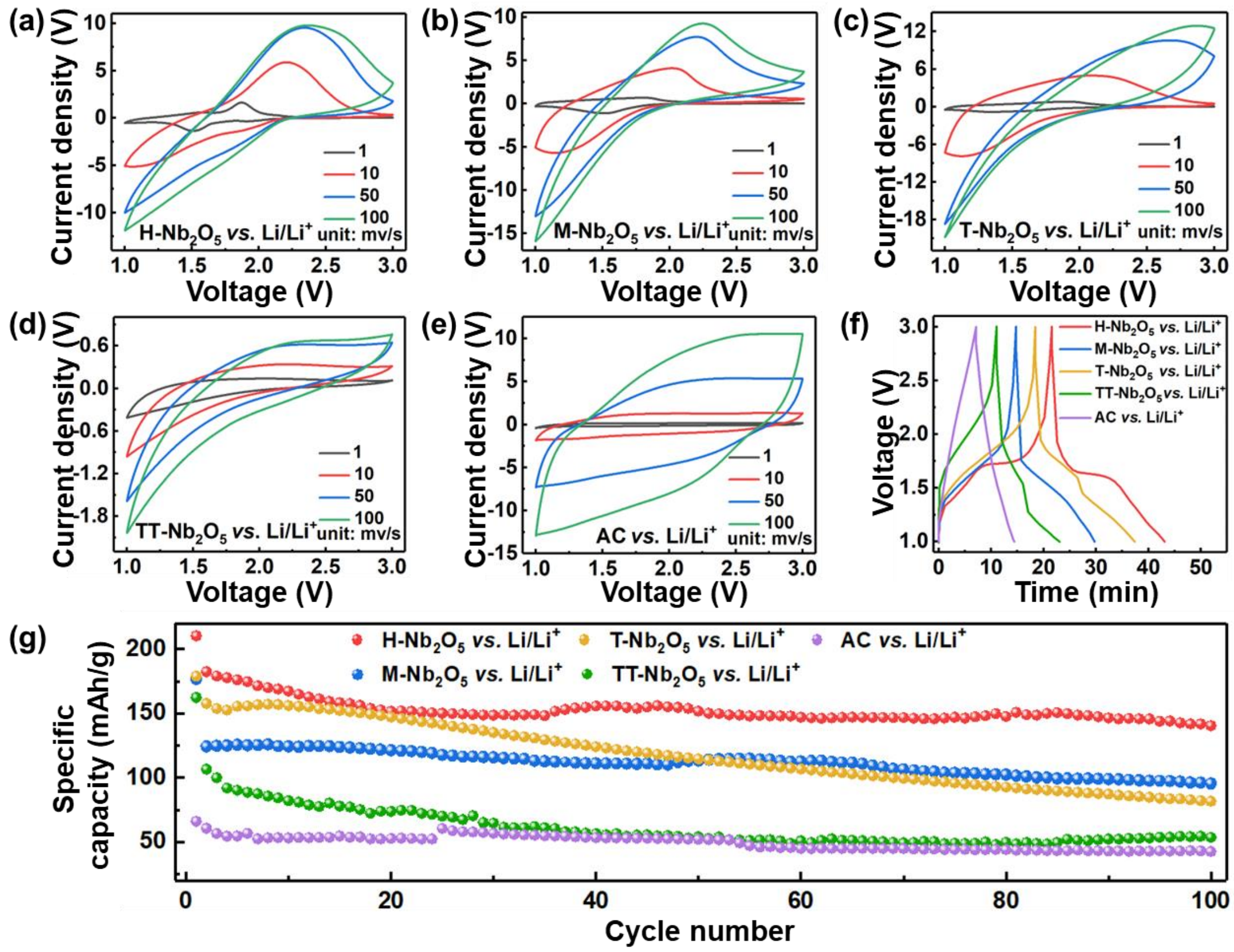

**Figure S7.** The half-cell electrochemical performance of the materials tested using lithium metal as the counter electrodes. (a-e) show the CV curves at different scan rates of H-$Nb_2O_5$, M-$Nb_2O_5$, T-$Nb_2O_5$, TT-$Nb_2O_5$ and AC, respectively. (f) gives the Galvanostatic charge/discharge profiles of the materials at 0.5 A/g. (g) gives the cycling performance.

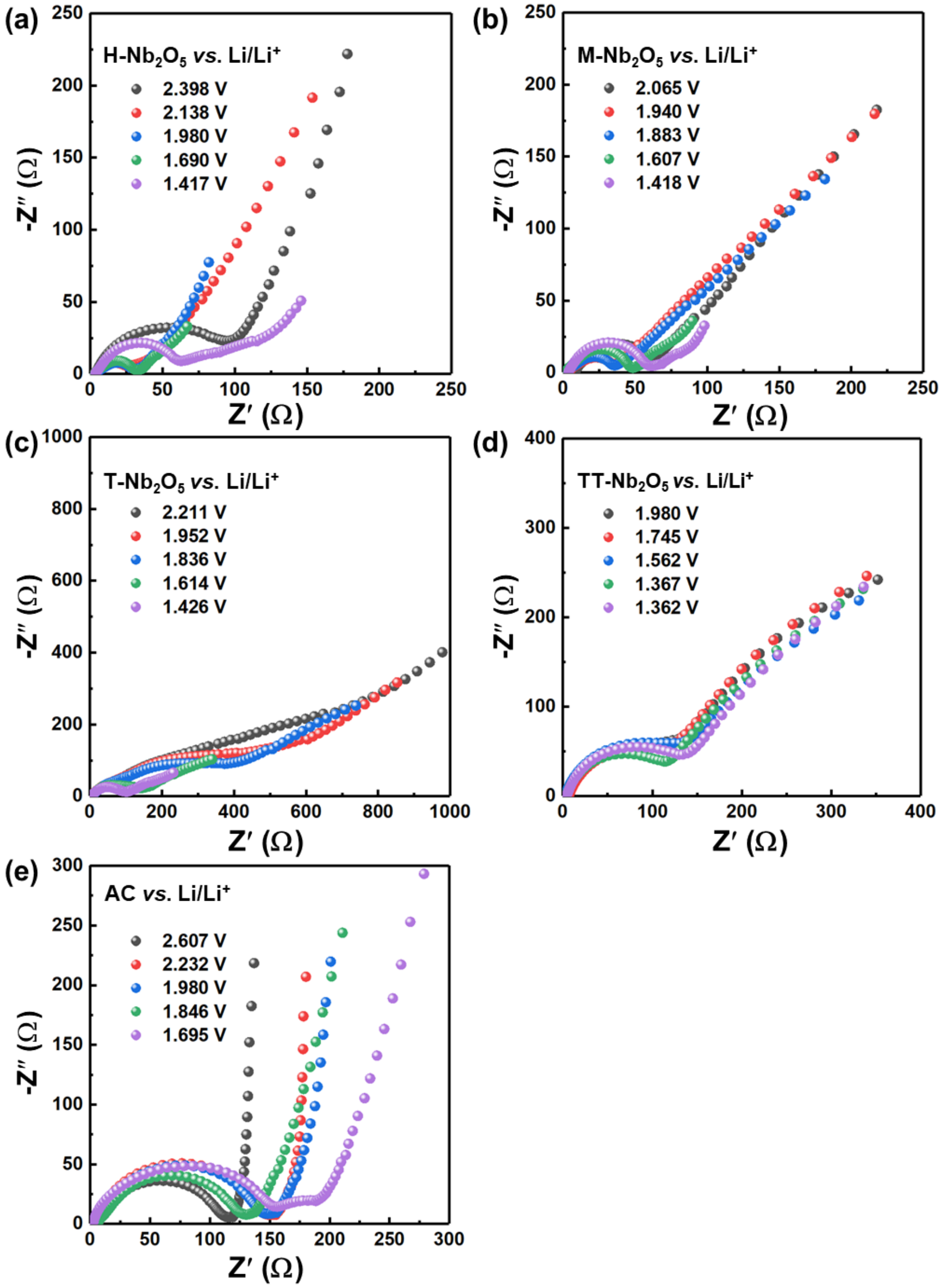

**Figure S8.** The EIS spectrum of H-$Nb_2O_5$ (a), M-$Nb_2O_5$ (b), T-$Nb_2O_5$ (c), TT-$Nb_2O_5$ (d) and AC (e) tested using lithium metal as the counter electrodes at different voltages.

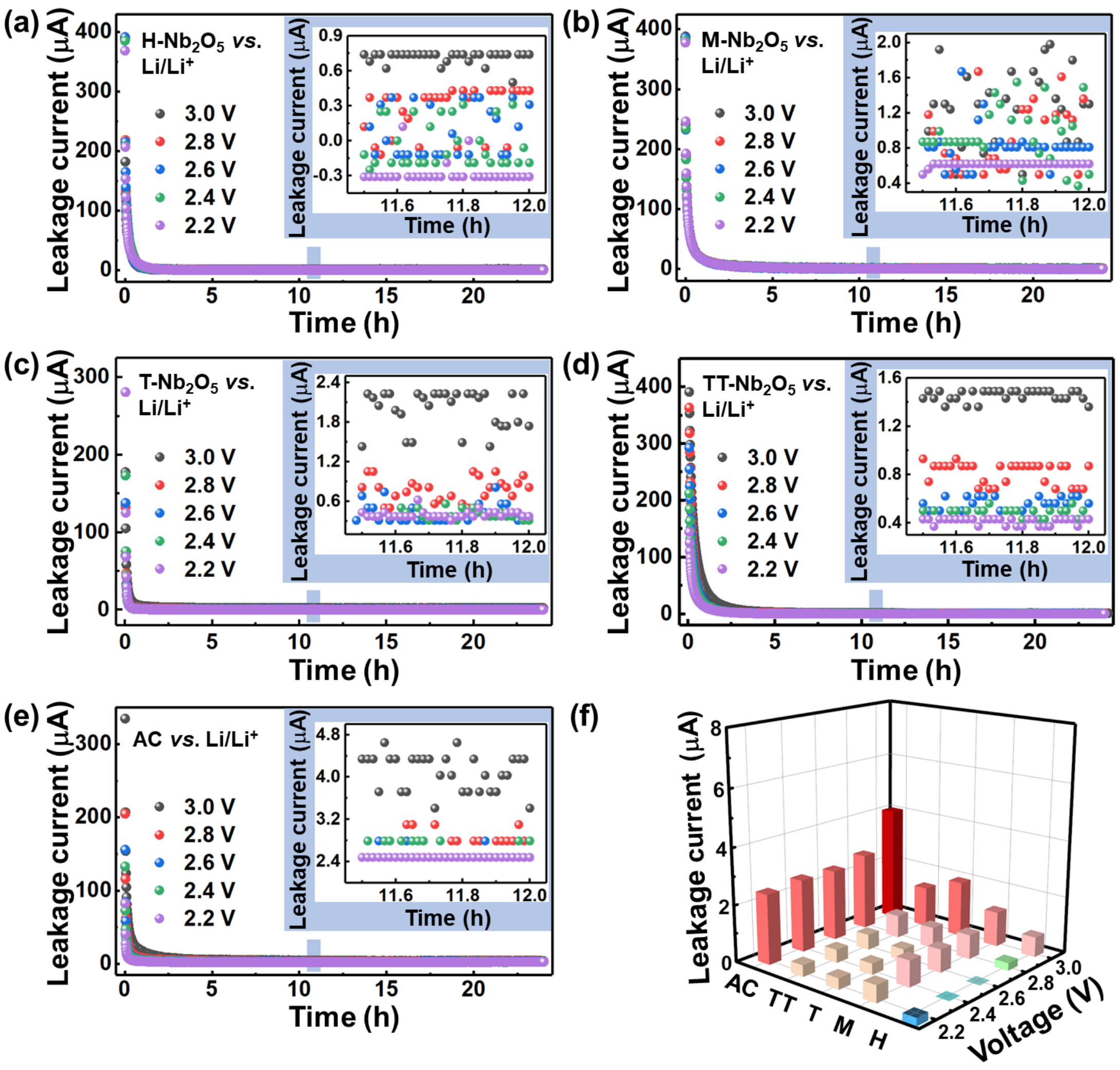

**Figure S9.** The leakage current of the materials tested using lithium metal as the counter electrodes. (a-e) show the current vs. time of H-$Nb_2O_5$ (a), M-$Nb_2O_5$ (b), T-$Nb_2O_5$ (c), TT-$Nb_2O_5$ (d) and AC (e) at different applied voltages, respectively. (f) compares the stabilized leakage current of the materials at various applied voltages.

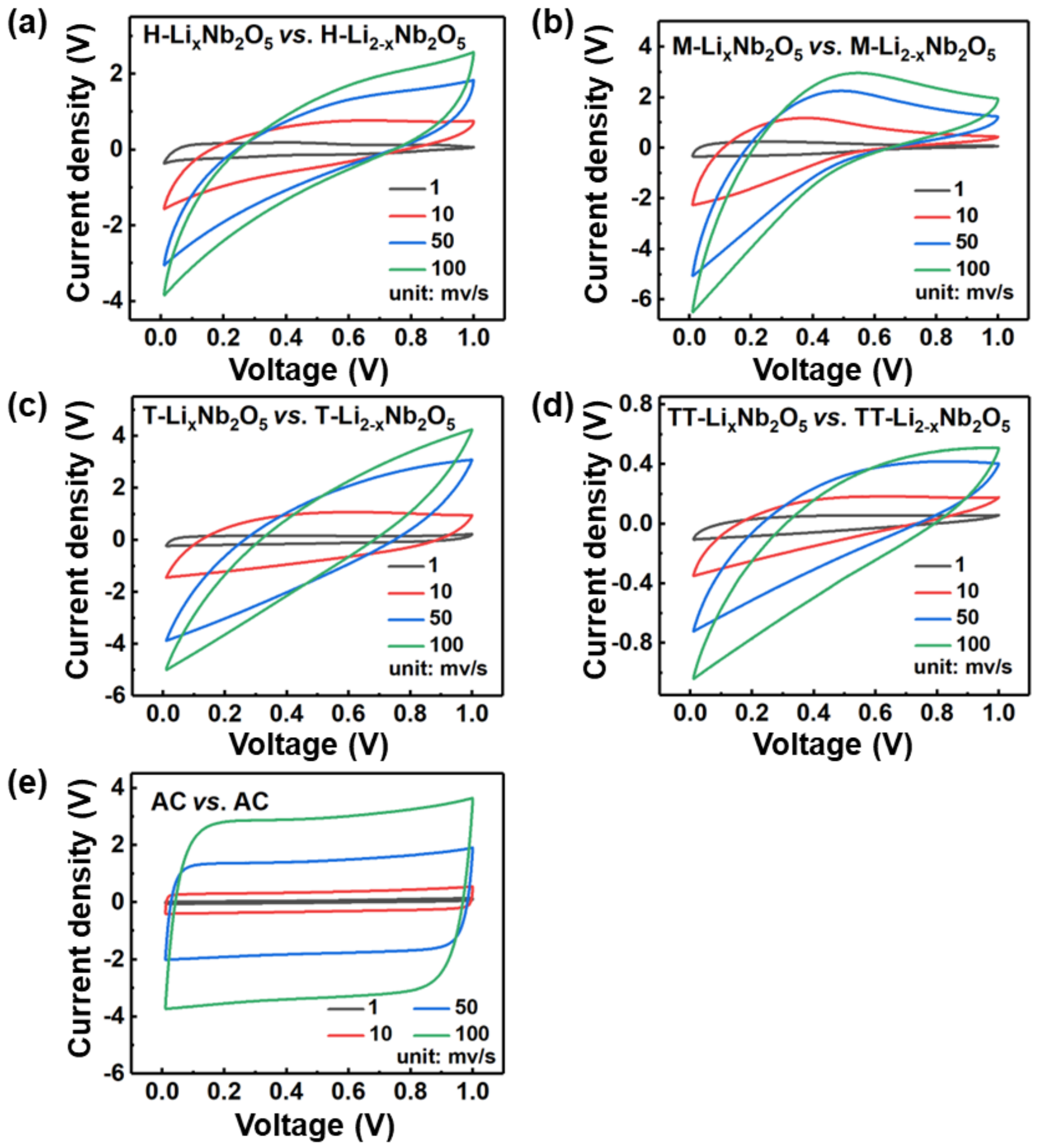

**Figure S10.** The CV curves of conjugately configured supercapacitors of H-$Nb_2O_5$ (a), M-$Nb_2O_5$ (b), T-$Nb_2O_5$ (c), TT-$Nb_2O_5$ (d) and AC (e) at various scan rates.

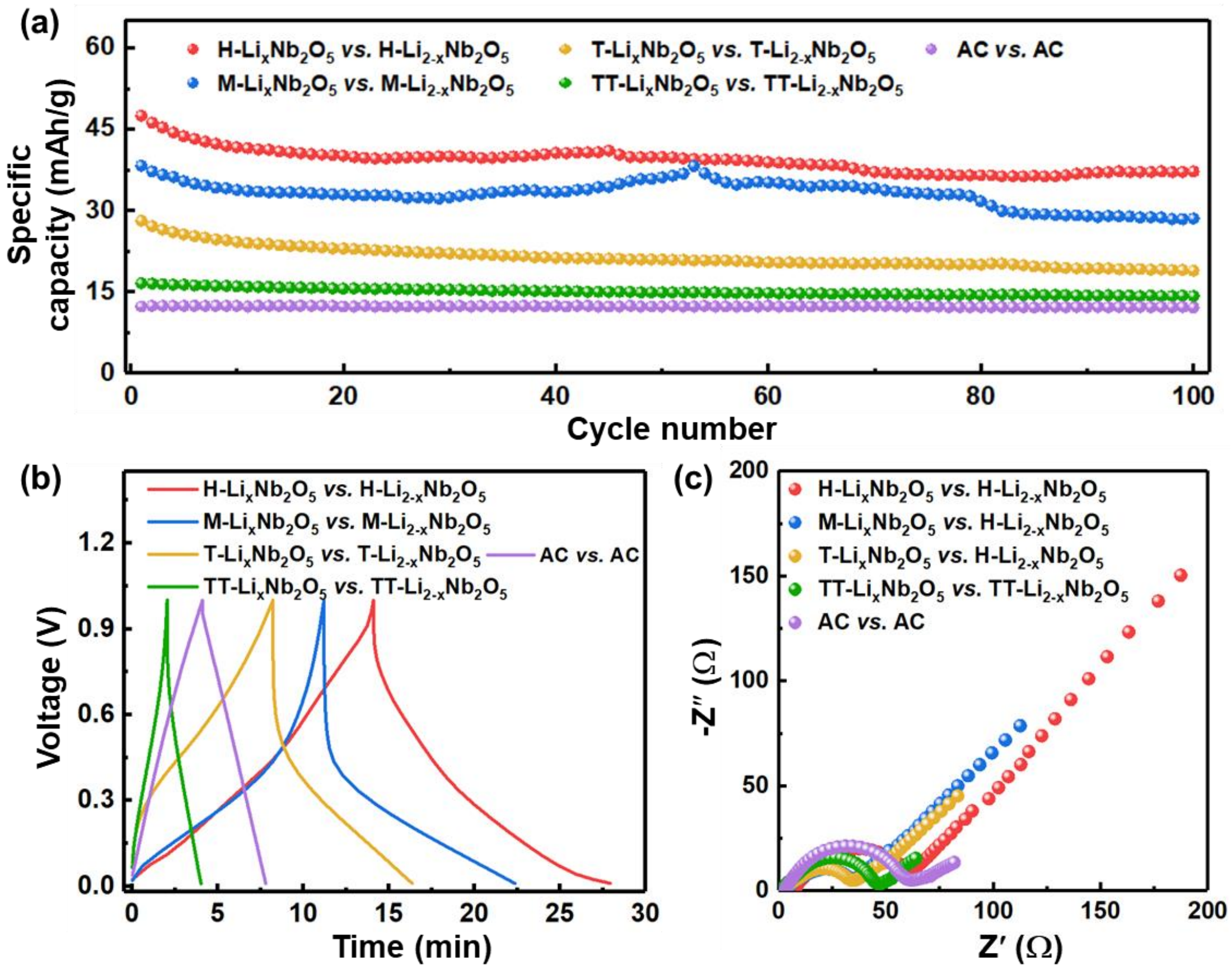

**Figure S11.** The electrochemical performance of the conjugately configured supercapacitors. (a) and (b) give the cycling performance and the Galvanostatic charge/discharge profiles of H-$Nb_2O_5$, M-$Nb_2O_5$, T-$Nb_2O_5$, TT-$Nb_2O_5$ and AC at the current density of 0.5 A/g, respectively. (c) gives the EIS measured at the 50% state-of-charge (0.5 V).

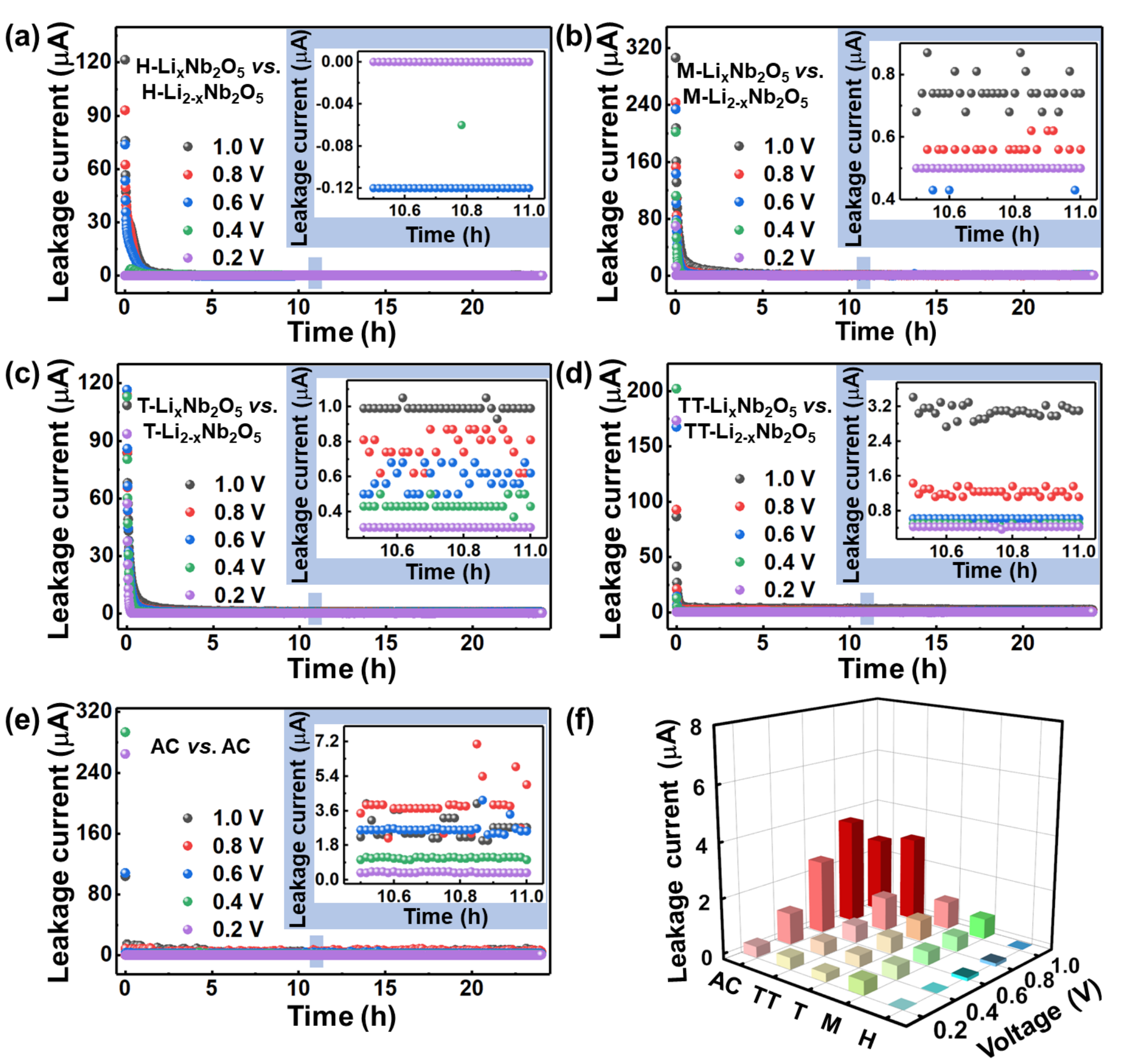

**Figure S12.** The leakage current of the conjugately configured supercapacitors. (a-e) show the current vs. time curves of H-$Nb_2O_5$ (a), M-$Nb_2O_5$ (b), T-$Nb_2O_5$ (c), TT-$Nb_2O_5$ (d) and AC (e) at different applied voltages, respectively. (f) compares the stabilized leakage current of the conjugately configured supercapacitors at various applied voltages.

| | H-$Nb_2O_5$ | M-$Nb_2O_5$ | T-$Nb_2O_5$ | TT-$Nb_2O_5$ |
|---|---|---|---|---|
| Site 1 | -917.372 | -1045.842 | -655.377 | -525.654 |
| Site 2 | -917.149 | -1045.821 | -655.000 | -525.663 |
| Site 3 | -917.006 | -1046.457 | -655.380 | -525.886 |
| Site 4 | -916.533 | -1046.692 | | -525.981 |
| Site 5 | -916.533 | -1046.711 | | -525.887 |
| Site 6 | -917.007 | -1046.455 | | -525.690 |
| Site 7 | -917.155 | -1045.826 | | -525.653 |
| Site 8 | -917.312 | -1045.824 | | |

**Table S1.** DFT calculated energy (eV) at every diffusion site of various $Nb_2O_5$ lattices.